\documentclass[letterpaper,final,10pt,onecolumn]{article}

\usepackage[dvips]{graphicx}

\usepackage{verbatim}
\usepackage{multirow}
\usepackage[square]{natbib}

\def\gsim{\;\lower4pt\hbox{${\buildrel\displaystyle >\over\sim}$}\;}
\def\lsim{\;\lower4pt\hbox{${\buildrel\displaystyle <\over\sim}$}\;}
\def\grls{\;\lower4pt\hbox{${\buildrel\displaystyle >\over <}$}\;}

\renewcommand{\vec}[1]{\mathbf{#1}}

\newcommand\addr[2]{{\footnotesize \it $^{#1}$#2}\\}

\begin{document}

\title{An Analytical Model Probing the Internal State of Coronal Mass
Ejections Based on Observations of Their Expansions and Propagations}

\author{Yuming Wang,$^{1,2}$ Jie Zhang,$^1$ and Chenglong Shen$^2$\\[1pt]
\addr{1}{Department of Computational and Data Sciences, George Mason University,}
\addr{}{Fairfax, VA 22030, USA; ymwang@ustc.edu.cn}\addr{2}{School of Earth \& Space Sci., Univ. of Sci. \& Tech. of
China, Hefei, Anhui 230026, China}}


\maketitle

\begin{abstract}
In this paper, a generic self-similar flux rope model is proposed to probe the
internal state of CMEs in order to understand the thermodynamic process and
expansion of CMEs in
interplanetary space. Using this model, three physical parameters and
their variations with heliocentric distance can be inferred based on
coronagraph observations of CMEs' propagation and expansion. One is
the polytropic index $\Gamma$ of the CME plasma, and the other two are
the average Lorentz force and the thermal pressure force
inside CMEs. By applying the model to the 2007
October 8 CME observed by STEREO/SECCHI, we find that
(1) the polytropic index of the CME plasma increased from initially 1.24
to more than 1.35 quickly, and then slowly decreased to about 1.34; it
suggests that there be continuously heat injected/converted into
the CME plasma and the value of $\Gamma$ tends to be $\frac{4}{3}$, a critical value
inferred from the model for a force-free flux rope;
(2) the Lorentz force directed inward while the thermal pressure force outward, and both
of them decreased rapidly as the CME moved out; the direction of
the two forces reveals that the thermal pressure force is the internal driver of the CME expansion
whereas the Lorentz force prevented the CME from expanding.
Some limitations of the model and approximations are discussed meanwhile.
\end{abstract}

\section{Introduction}
Coronal mass ejections (CMEs) are the most energetic eruptive phenomenon
occurring in the Sun's atmosphere and the major driver of space weather. They
carry a huge amount of mass, kinetic energy and magnetic flux into the interplanetary space,
and therefore may cause many significant consequences in the geospace.
In this paper, we develop a generic flux rope model to infer three physical
parameters of CMEs as well as their variations with heliocentric distance through the usage of  the latest STEREO (Solar
TErrestrial RElations Observatory) observations. The first parameter,
is the polytropic index, $\Gamma$, that describes the thermodynamic
process of the gas, and the other two are the Lorentz force and
the thermal pressure force, that reveal the internal cause of
the CME expansion.

CMEs have been observed and studied for decades. There have been many observations, either
through remote sensing observations or in-situ samplings, revealing the internal properties of CME plasmas.
For example, the remote sensing data from SOHO/UVCS (ultraviolet
coronagraph spectrometers, \citet{Kohl_etal_2006}) can diagnose the plasma temperature, density, velocity and heating at a few solar radii from the Sun. Such
spectroscopic analyses suggested that CMEs be a loop-like structure
\citep[e.g.,][]{Ciaravella_etal_2003} with helical magnetic field \citep[e.g.,][]{Antonucci_etal_1997,
Ciaravella_etal_2000}, and probably have a higher temperature than that in the typical
Corona in the near Sun region \citep{Ciaravella_etal_2003}. The thermal energy
deposited into CME plasmas is roughly comparable to the kinetic and gravitational
potential energies of CMEs in the inner corona \citep[e.g.,][]{Akmal_etal_2001, Ciaravella_etal_2001}.
Some Internal properties of CMEs can also be revealed from in-situ observations, e.g, by Ulysses and ACE spacecraft. For example, the interplanetary CMEs (ICMEs) at 1 AU usually show
a lower temperature and stronger magnetic fields than that in the ambient solar wind
\citep[e.g.,][]{Burlaga_etal_1981, Richardson_Cane_1995, Farrugia_etal_1993b}. The ion
charge states in ICMEs are often higher \citep[e.g.,][]{Lepri_etal_2001,
Lynch_etal_2003}. Based on the analysis of ion charge states, it was also found  that the thermal energy input to the CME plasmas is
at the same order of the CME kinetic energy \citep[e.g.,][]{Rakowski_etal_2007}.

The above studies provided the information of the internal properties of
CMEs, but only at a certain position and/or at a certain time. What is  largely lacking is the global observations and thus the global understanding  of the evolution of the internal state of CMEs during their continuous propagation
throughout the interplanetary medium. What thermodynamic process does the CME plasma undergo?
What happens with the various forces involved?  Limited knowledge on these global issues were obtained through  indirect ways, largely from the statistical
combination of observations of many CMEs from multiple spacecraft over a long time period. The multiple-spacecraft
measurements suggested that the polytropic index of CME plasmas be  below 1.3
\citep[e.g.,][]{Liu_etal_2006}, which is apparently different from that of solar wind, which is
about 1.46 \citep{Totten_etal_1995}. The radial widths of CMEs continuously increase at the
order of local Alfv\'{e}n speed as they move away from the Sun within 10 AU \citep[e.g.,][]{Wang_Richardson_2004,
Wang_etal_2005a, Jian_etal_2008}, and the magnetic field decreases faster in ICMEs
than in ambient solar wind but the temperature does not \citep[e.g.,][]{Wang_Richardson_2004,
Wang_etal_2005a, Liu_etal_2006}. It should be noted that all the above conclusions
were established on statistical surveys, which can  not  review the detailed evolution behavior
of any individual CMEs.

It is now well accepted that CMEs, at least a significant percentage of them, have a flux
rope-like structure (Fig.\ref{fg_coor}). Thus we try to study the internal state of CMEs by establishing a flux rope model.
 There are already various flux rope models concerning CME initiation and/or propagation \citep[e.g.,][]{Burlaga_etal_1981,
Goldstein_1983, Chen_1989, Forbes_Isenberg_1991, Kumar_Rust_1996, Vandas_etal_1997a,
Gibson_Low_1998, Titov_Demoulin_1999}. These models have their own specific purposes, and may not suit
the issues attacked in this paper. We present our model in the next section.  We then make a case study in
Sec.\ref{sec_case} by applying the model to the CME that occurred on 2007 October 8, whose
expansion and propagation over a large distance throughout the interplanetary space were well observed.
In Sec.\ref{sec_summary}, a brief summary is given. Finally, we thoroughly discuss the limitations
and approximations of the model in Sec.\ref{sec_discussion}.

\section{The Model}\label{sec_model}
\subsection{Derivation and Parameters}
\begin{figure*}[tbh]
  \centering
  \includegraphics[width=.6\hsize]{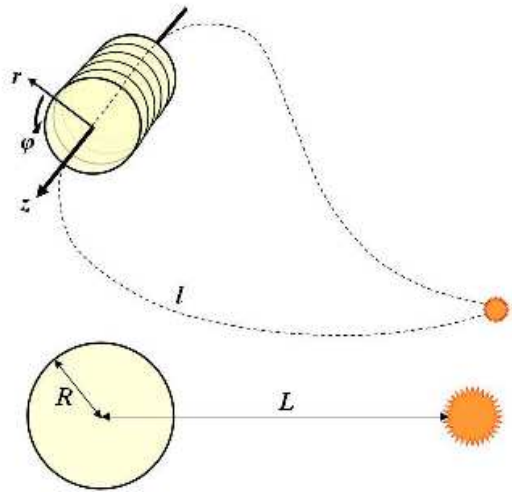}
  \caption{{\it Upper panel:} Schematic plot of a flux-rope CME.
{\it Lower panel:} The cross-section of the flux rope. $R$ is the
radius, $L$ is the distance from the flux rope axis to the solar surface,
and $l$ is the axial length of the flux rope. The distance, $h$,
of the CME leading edge and the minimum
and the maximum position angles (indicated by the dot-dashed lines) are
the three directly measurable quantities from  coronagraph observations.}\label{fg_coor}
\end{figure*}

As commonly assumed, CMEs are approximated as a cylindrical flux rope in a
local scale, even though in the global scale they may be a loop-like structure
with two ends rooted on the surface of the Sun. The flux rope can therefore
be treated axisymmetrically in the cylindrical coordinates ($r$, $\phi$, $z$)
with the origin on the axis (ref to Fig.\ref{fg_coor}) and
$\frac{\partial}{\partial\phi}=\frac{\partial}{\partial z}=0$. As a result,
only a 2-dimensional circular cross-section of the flux rope needs to be
considered in our model. Let the radius of the circular cross-section be $R$,
the expansion speed of the flux rope is given by
\begin{eqnarray}
 v_e(t)=\frac{dR(t)}{dt}
\end{eqnarray}
Further, assuming that the flux rope is undergoing a self-similar expansion within the cross section,
we can get a dimensionless variable
\begin{eqnarray}
 x=\frac{r}{R} \label{eq_rrx}
\end{eqnarray}
which is the normalized radial distance from the flux rope axis and
independent of time. The $x=1$ is the boundary of the flux rope. Therefore,
the $r$-component of the velocity is
\begin{eqnarray}
 v_r(t, x)=\frac{d r}{d t}=xv_e(t)  \label{eq_vr}
\end{eqnarray}
and the acceleration is
\begin{eqnarray}
a_r(t,x)&=&\frac{d\vec v(t,x)}{dt}\cdot\hat{\vec
r}=xa_e(t)-\frac{v_\phi^2(t,x)}{xR}\label{eq_ar}
\end{eqnarray}
where
\begin{eqnarray}
a_e(t)=\frac{dv_e(t)}{dt}
\end{eqnarray}
is the acceleration of the expansion. The first term on the
right-hand side of Eq.\ref{eq_ar} is the acceleration of the radial
motion of the plasma, and the second term is the acceleration
contributed from the circular/poloidal motion.

With the above preliminary preparation, we now investigate an arbitrary fluid
element in the flux rope by starting from the momentum conservation equation
(in the frame frozen-in with the moving flux rope),
\begin{eqnarray}
 \rho\frac{\partial\vec{v}}{\partial t}+\rho(\vec{v}\cdot\nabla)\vec{v}=-\nabla p+\vec{j}\times\vec{B}\label{eq_mom_1}
\end{eqnarray}
where $\rho$ is the density, $p$ is the plasma thermal pressure,
$\vec{B}=(0,B_\phi,B_z)$ is the magnetic induction, and
$\vec{j}=\frac{1}{\mu_0}\nabla\times\vec{B}$ is the current density.
Here, the viscous stress tensor, $S$, gravity, $\vec{F_g}$, and the
equivalent force due to the use of a non-inertial reference frame,
$\vec{F_a}$ opposite to the acceleration, are ignored (the
validation of this treatment will be discussed at the end of this
paper). This equation can be decomposed into the $\hat{r}$ and
$\hat{\phi}$ components as follows
\begin{eqnarray}
\hat{r}: &&\rho\frac{\partial v_r}{\partial t}+\rho(v_r\frac{\partial v_r}{\partial r}-\frac{v_\phi^2}{r})=-\frac{\partial p}{\partial r}+(\vec{j}\times\vec{B})_r\label{eq_mom_2}  \\
\hat{\phi}: &&\rho\frac{\partial v_\phi}{\partial t}+\rho(v_r\frac{\partial v_\phi}{\partial r}+\frac{v_rv_\phi}{r})=0\label{eq_mom_3}
\end{eqnarray}
According to the self-similar assumption, $v_\phi$ has the following
form
\begin{eqnarray}
v_\phi(t, x)=f_p(x)v_p(t)
\end{eqnarray}
in which $f_p$ is a function of only $x$ and $v_p$ is a function of only $t$.
Combine it with Eq.\ref{eq_vr} and \ref{eq_mom_3}, it is inferred that
\begin{eqnarray}
v_\phi=k_1x^{k_2-1}R^{-k_2} \label{eq_vphi}
\end{eqnarray}
where $k_1$ and $k_2$ are both constants. It is required that
$k_2\geq1$ to guarantee that $v_\phi$ is physically
meaningful, $k_2=1$ implies that the angular momentum of the flux rope
is conserved, and $k_2>1$ means that the angular momentum decreases as
the flux rope expands. Combine Eq.\ref{eq_rrx}, \ref{eq_vr}, \ref{eq_mom_2} and
\ref{eq_vphi}, we can rewrite the momentum conservation equation
with $\hat{r}$ as
\begin{eqnarray}
(\vec{j}\times\vec{B})_r=\rho(a_ex-k_1^2R^{-2k_2-1}x^{2k_2-3})+R^{-1}\frac{\partial p}{\partial x}
\label{eq_mom_4}
\end{eqnarray}
For a thermodynamic process, we relate the thermal
pressure $p$ with the density $\rho$ by the polytropic equation of state
\begin{eqnarray}
 p=k_3\rho^\Gamma \label{eq_prho}
\end{eqnarray}
where $k_3$ is a positive constant and $\Gamma$ is a variable treated as
the polytropic index, and Eq.\ref{eq_mom_4} becomes
\begin{eqnarray}
(\vec{j}\times\vec{B})_r=\rho(a_ex-k_1^2R^{-2k_2-1}x^{2k_2-3})+k_3R^{-1}\frac{\partial \rho^\Gamma}{\partial x} \label{eq_mom_5}
\end{eqnarray}
Define a quantity $f_{em}$ to be the average Lorentz force
over $\hat{r}$ from the axis to the boundary of the flux rope,
$f_{em}=\frac{1}{R}\int_0^R(\vec{j}\times\vec{B})_rdr$.
From Eq.\ref{eq_mom_5}, we get
\begin{eqnarray}
f_{em}=a_e\int_0^1\rho xdx-k_1^2R^{-2k_2-1}\int_0^1\rho x^{2k_2-3}dx+k_3R^{-1}\int_0^1\frac{\partial\rho^\Gamma}{\partial x}dx \label{eq_pem1}
\end{eqnarray}
$f_{em}>0$ means that the average Lorentz force directs outward from the
axis of the flux rope, causing expansion. On the other hand, $f_{em}<0$
prevents the expansion of the flux rope.

We assume that the mass of a CME is conserved when it propagates in the
outer corona and interplanetary space, where the CME has fully developed.
The mass conservation gives
\begin{eqnarray}
 \int\rho rdr d\phi dz=2\pi lR^2\int_0^1\rho xdx=M \mathrm{\ (constant)} \label{eq_rho}
\end{eqnarray}
where $l$ is the axial length of the flux rope (Fig.\ref{fg_coor}).
Since the flux rope is assumed to be self-similar and it is generally
accepted that the magnetic field lines are frozen-in with the plasma flows
in corona/interplanetary space, the density in the flux rope has a fixed
distribution $f_\rho(x)$, and therefore
\begin{eqnarray}
\rho(t, x)=f_\rho(x)\rho_0(t)
\end{eqnarray}
Define positive constants
\begin{eqnarray}
k_4=\int_0^1f_\rho xdx \\
k_5=\int_0^1f_\rho x^{2k_2-3}dx
\end{eqnarray}
and a variable
\begin{eqnarray}
q(\Gamma)=f_\rho^\Gamma(0)-f_\rho^\Gamma(1)
\end{eqnarray}
Then it can be inferred from Eq.\ref{eq_rho} that
\begin{eqnarray}
\rho_0=\frac{1}{2\pi}k_{4}^{-1}MR^{-2}l^{-1}\label{eq_rho_3}
\end{eqnarray}
and Eq.\ref{eq_pem1} can be written as
\begin{eqnarray}
f_{em}=\frac{M}{2\pi}(a_eR^{-2}l^{-1}-k_1^2k_4^{-1}k_5R^{-2k_2-3}l^{-1})-f_{th} \label{eq_pem2}
\end{eqnarray}
where
\begin{eqnarray}
f_{th}=\frac{1}{R}\int_0^R-\frac{\partial p}{\partial r}dr=k_3q\rho_0^\Gamma R^{-1} \label{eq_pth}
\end{eqnarray}
is the average thermal pressure force. Like $f_{em}$, $f_{th}$ points outward
if it is larger than zero.

On the other hand, in an axisymmetric cylindrical flux rope,
\begin{eqnarray}
&\vec{B}=B_\phi\hat{\phi}+B_z\hat{z}=\nabla\times\vec{A} \label{eq_bcomp} \\
&B_\phi=-\frac{\partial A_z}{\partial r} \label{eq_b_phi} \\
&B_z=\frac{1}{r}\frac{\partial}{\partial r}(rA_\phi) \label{eq_b_z} \end{eqnarray}
As the magnetic flux is conserved in both $\hat\phi$ and $\hat{z}$ directions,
we get
\begin{eqnarray}
&\Phi_\phi=-l\int_0^R\frac{\partial A_z}{\partial r} dr=l(A_z(0)-A_z(R)) \label{eq_Phi_phi} \\
&\Phi_z=2\pi\int_0^R\frac{\partial}{\partial r}(rA_\phi) dr=2\pi RA_\phi(R) \label{eq_Phi_z}
\end{eqnarray}
In order to satisfy the self-similar expansion assumption, $A_\phi$ and $A_z$
have to keep their own distributions, respectively. Thus, according to the above two
equations,
\begin{eqnarray}
 A_\phi(t, x)&=&\frac{f_\phi(x)}{R} \label{eq_A_phi} \\
 A_z(t, x)&=&\frac{f_z(x)}{l} \label{eq_A_z}
\end{eqnarray}
It can be proved that the conservation of helicity is satisfied automatically
\begin{eqnarray}
H_m&=&\int\vec{B}\cdot\vec{A}rdrd\phi dz \nonumber \\
&=&2\pi \int_0^1 [\frac{f_z}{x}\frac{\partial}{\partial x}(xf_\phi)-f_\phi\frac{\partial f_z}{\partial x}] xdx \nonumber \\
&=&\mathrm{constant}
\end{eqnarray}
Combining Eq.\ref{eq_b_phi}, \ref{eq_b_z}, \ref{eq_A_phi}
and \ref{eq_A_z}, we can calculate the Lorentz force in the flux
rope
\begin{eqnarray}
\vec{j}\times\vec{B}&=&\frac{1}{\mu_0}(\nabla\times\vec{B})\times\vec{B} \nonumber \\
&=&-\mu_0^{-1}R^{-5}\{x^{-2}\frac{\partial}{\partial x}(xf_\phi)\frac{\partial^2}{\partial x^2}(xf_\phi)-x^{-3}[\frac{\partial}{\partial x}(xf_\phi)]^2\}\hat{r} \nonumber\\
&&-\mu_0^{-1}R^{-3}l^{-2}x^{-1}\frac{\partial f_z}{\partial x}\frac{\partial}{\partial x}(x\frac{\partial f_z}{\partial x})\hat{r} \label{eq_jb_3}
\end{eqnarray}
and therefore
\begin{eqnarray}
f_{em}&=&-\mu_0^{-1}R^{-5}\int_0^1\{x^{-2}\frac{\partial}{\partial x}(xf_\phi)\frac{\partial^2}{\partial x^2}(xf_\phi)-x^{-3}[\frac{\partial}{\partial x}(xf_\phi)]^2\}dx \nonumber\\
&&-\mu_0^{-1}R^{-3}l^{-2}\int_0^1x^{-1}\frac{\partial f_z}{\partial x}\frac{\partial}{\partial x}(x\frac{\partial f_z}{\partial x})dx \label{eq_jbr_0} \nonumber \\
&=&-\mu_0^{-1}k_6R^{-5}-\mu_0^{-1}k_7R^{-3}l^{-2} \label{eq_pem3}
\end{eqnarray}
where $k_6$ and $k_7$ are both constants. It could be proved that
the sign of $k_6$ is determined by $B_z^2(R)-B_z^2(0)$, and $k_7\geq0$.

The two forms of $f_{em}$, Eq.\ref{eq_pem2} and \ref{eq_pem3}, result in
\begin{eqnarray}
&a_e-k_1^2k_4^{-1}k_5R^{-2k_2-1} \nonumber\\
&=-2\pi\mu_0^{-1}M^{-1}(k_6R^{-3}l+k_7R^{-1}l^{-1})+2\pi M^{-1}k_3(2\pi k_4M^{-1}R^2l)^{-\Gamma}[f_\rho^\Gamma(0)-f_\rho^\Gamma(1)]Rl \label{eq_acs}
\end{eqnarray}
in which $f_{th}$ is substituted by Eq.\ref{eq_pth}.
As at present it is impossible to practically detect the axial length of a
flux rope, here we will relate it with a measurable variable, $L$, the distance
between the flux rope axis and the solar surface (Fig.\ref{fg_coor}),
at which altitude the flux rope originates, by the assumption
\begin{eqnarray}
l=k_8L \label{eq_ll}
\end{eqnarray}
where $k_8$ is a positive constant. The topology of
flux rope as shown in Figure \ref{fg_coor} implies that this
assumption is reasonable.
Finally, Eq.\ref{eq_acs} can be simplified to
\begin{eqnarray}
a_e-c_0R^{-c_1-3}=-c_2R^{-3}L-c_3R^{-1}L^{-1}+c_4(c_5^\Gamma-c_6^\Gamma)R^{1-2\Gamma}L^{1-\Gamma} \label{eq_sase_fitting}
\end{eqnarray}
where
\begin{eqnarray}
&c_0=k_1^2k_4^{-1}k_5\geq 0 \\
&c_1=2k_2-2 \geq 0 \\
&c_2=2\pi\mu_0^{-1}M^{-1}k_6k_8 \\
&c_3=2\pi\mu_0^{-1}M^{-1}k_7k_8^{-1}\geq 0 \\
&c_4=2\pi M^{-1}k_3k_8>0 \\
&c_5=(2\pi)^{-1}Mk_4^{-1}k_8^{-1}f_\rho(0) \geq0 \label{eq_density_center} \\
&c_6=(2\pi)^{-1}Mk_4^{-1}k_8^{-1}f_\rho(1) \geq0 \label{eq_density_boundary}
\end{eqnarray}
The left-hand side of Eq.\ref{eq_sase_fitting} describes the motion of the fluids
in the flux rope. Its first item is the acceleration due to the radial motion (i.e., expansion)
and the second one gives the acceleration due to the poloidal motion. The right-hand
side reflects the contributions from the Lorentz force (the first two items) and
thermal pressure force (the last one).
The constants $k_{1-8}$ and $c_{0-6}$ appeared above are summarized in Table \ref{tb_constants}.

\begin{table*}[t]
\caption{List of the constants $k_{1-8}$ and $c_{0-6}$}
\label{tb_constants}
\begin{tabular}{cp{120pt}|cp{120pt}}
\hline
Constant & Interpretation & Constant & Interpretation \\
\hline
$k_1$ & Scale the initial magnitude of the poloidal motion &$k_4$ and $k_5$ & Integral constants related to the density distribution \\
$k_2$ & Decrease rate of the angular momentum as the flux rope expands &$k_6$ and $k_7$ & Scale the initial magnitude of the Lorentz force contributed by the axial and poloidal fields \\
$k_3$ & Coefficient in the polytropic equation of state &$k_8$ & Assumed constant to relate the length of flux rope $l$ to distance $L$ \\
\hline\hline
$c_0$ & Scale the initial magnitude of the acceleration due to the poloidal motion &$c_2$ and $c_3$ & Similar to $k_6$ and $k_7$ \\
$c_1$ & Similar to $k_2$ &$c_4(c_5^\Gamma-c_6^\Gamma)$ & Scale the initial magnitude of the contribution by thermal pressure force \\
\hline
\end{tabular}
\end{table*}

The Lorentz force and thermal pressure force can be rewritten in terms of the
constants $c_{0-6}$, $k_8$ and the total mass $M$ as follows
\begin{eqnarray}
f_{em}=-\frac{M}{2\pi k_8}(c_2R^{-5}+c_3R^{-3}L^{-2})\label{eq_fem}\\
f_{th}=\frac{M}{2\pi k_8}c_4(c_5^\Gamma-c_6^\Gamma)R^{-2\Gamma-1}L^{-\Gamma}\label{eq_fth}
\end{eqnarray}
and their ratio is
\begin{eqnarray}
\frac{f_{em}}{f_{th}}=\frac{c_2R^{2\Gamma-4}L^\Gamma+c_3R^{2\Gamma-2}L^{\Gamma-2}}{c_4(c_6^\Gamma-c_5^\Gamma)}\label{eq_inff}
\end{eqnarray}

In summary, starting from MHD equations with the three major assumptions that
(1) the flux-rope CME has an axisymmetric cylinder configuration, (2) its cross-section is
self-similarly evolving, and (3) its axial length is proportional to the distance
from the solar surface, we find that the polytropic index, $\Gamma$, can be related to the
measurable parameters: the distance, $L$, the radius, $R$, and another derived
quantity, the expansion acceleration ($a_e$),
as shown in Eq.\ref{eq_sase_fitting}.
If we have enough measurement points, the unknown constants $c_{0-6}$ and variable
$\Gamma$ could be obtained through some fitting techniques (e.g., that described in the first
paragraph of Sec.\ref{sec_results}), and then the relative strength
of the Lorentz force and thermal pressure force can also be easily calculated by Eq.\ref{eq_fem} and \ref{eq_fth}.

\subsection{Asymptotic Value of Polytropic Index $\Gamma$} \label{sec_asymptotic}
Here, we consider the case of a nearly force-free expanding flux rope.
It is generally true that most CMEs are almost force-free at least near 1 AU though
they may be far away from a froce-free state at initial stage.
It can be proved that $R\propto L$
(ref. to Appendix), i.e., $R=\alpha L$ where $\alpha$ is a positive constant.
Then Eq.\ref{eq_inff} becomes
\begin{eqnarray}
\frac{f_{em}}{f_{th}}=\frac{(c_2\alpha^{2\Gamma-4}+c_3\alpha^{2\Gamma-2})L^{3\Gamma-4}}{c_4(c_6^\Gamma-c_5^\Gamma)}
\end{eqnarray}
It is found that $\Gamma=\frac{4}{3}$ is a critical point, above/below which
the absolute value of Lorentz force decreases slower/faster than that of thermal
pressure force as increasing distance $L$. This value of $\Gamma$ is the same as
that obtained by \citet{Low_1982} and \citet{Kumar_Rust_1996} for a self-similar
expanding flux rope. This inference is reasonable because smaller $\Gamma$ implies the
plasma absorbs more heat for the same expansion and therefore the thermal pressure
should decrease slower.

Under force-free condition, Eq.\ref{eq_sase_fitting} can also reduce to
\begin{eqnarray}
a_e=c_0(\alpha L)^{-c_1-3}-c_2\alpha^{-3}L^{-2}-c_3\alpha^{-1}L^{-2}+c_4(c_5^\Gamma-c_6^\Gamma)\alpha^{1-2\Gamma}L^{2-3\Gamma}
\end{eqnarray}
and at infinite distance, $L\rightarrow+\infty$, we have
\begin{eqnarray}
a_{e\infty}\sim c_4(c_5^{\Gamma_\infty}-c_6^{\Gamma_\infty})\alpha^{1-2{\Gamma_\infty}}L_\infty^{2-3{\Gamma_\infty}}
\end{eqnarray}
The above equation indicates that $\Gamma=\frac{2}{3}$ is another critical point.
The polytropic index $\Gamma$ should be larger than $\frac{2}{3}$ to make sure that the flux rope will finally approach a
steady expansion and propagation state (including the case that the flux rope stop somewhere
without expansion). Otherwise, the flux rope will always accelerated expanding.

Based on the current observations, the expansion behavior of CMEs at large heliocentric
distance is not as clear as that in the inner heliosphere. The investigations on
the radial widths of CMEs suggest that CMEs at least keep expanding within about 15 AU
\citep[e.g.,][]{Wang_Richardson_2004, Wang_etal_2005a}, but the expansion speeds seem
to be slower and slower. Although the number of CMEs observed near and beyond 15 AU is
small and the uncertainty of statistics is large, it is likely that a CME may not be
able to keep an accelerated expansion always. Thus, in practice, the polytropic index
of the CME plasma should be larger than $\frac{2}{3}$.

\section{The 2007 October 8 CME} \label{sec_case}
\subsection{Observations and Measurements} \label{sec_measurements}
The suite of SECCHI instruments on board STEREO spacecraft provide an
unprecedented continuous view of CMEs from the surface of the Sun through
the inner heliosphere. The instruments, EUVI, COR1, COR2, HI1 and HI2,
make the images of the solar corona in the ranges of 0--1.5, 1.4--4.0,
2.5--15.0, $\sim$15--90, and $\sim$90--300 solar radii ($R_S$),
respectively \citep{Howard_etal_2008}. The SECCHI observations present
us the great opportunity to study the evolution of CMEs over an extended
distance. The CME launched on 2007 October 8 is a well observed event,
which is used to study the CME evolution and the applicability of our model.

\begin{figure*}[p]
  \centering
  \includegraphics[width=0.9\hsize]{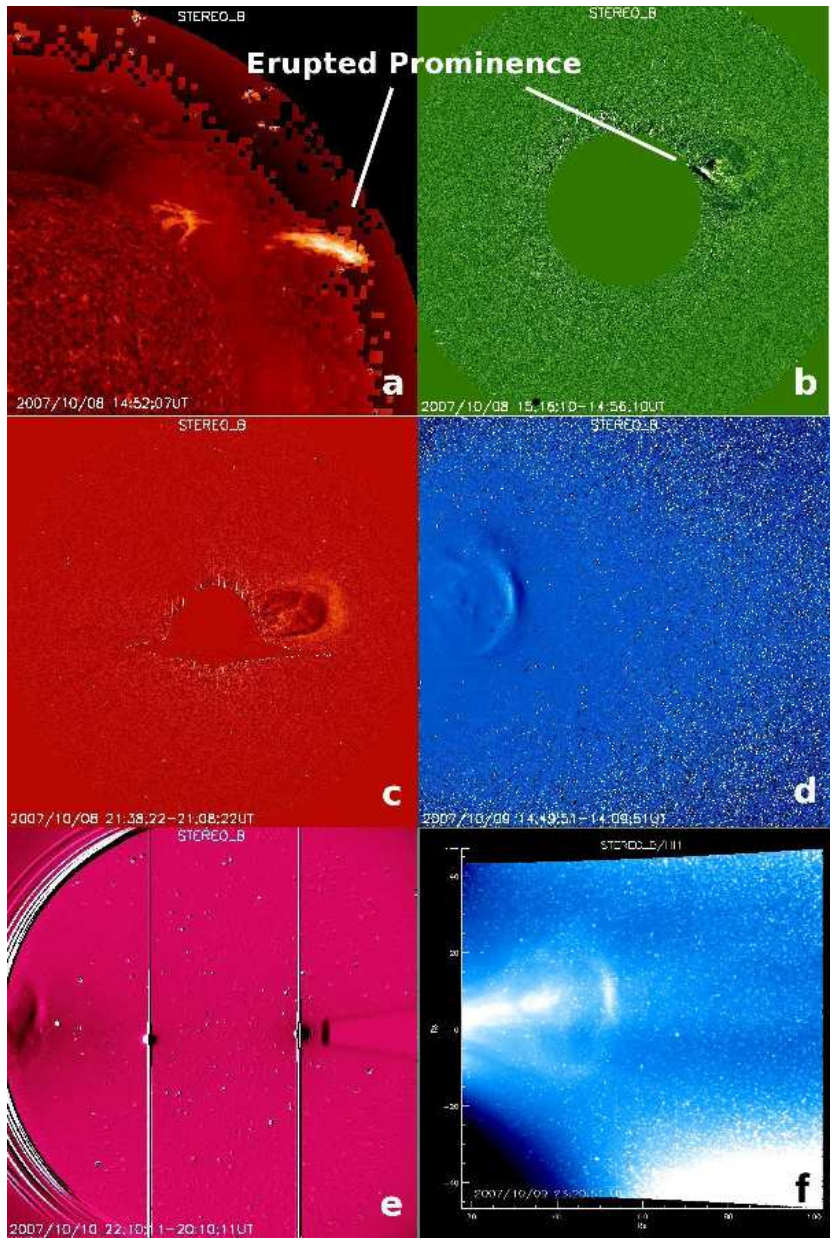}
  \caption{Images of the 2007 October 8 CME taken by (a) EUVI at 304\AA,
(b) COR1, (c) COR2, (d) HI1 and (e) HI2 on board STEREO B. The panel
at lower-right corner shows a direct image of the CME in the HI1
FOV. This image has been corrected to the plane perpendicular to the
line between the Sun and STEREO B, because the CME is assumed to be
a limb event in CORs and the direction that HI1 camera faces to is
different with CORs'.}\label{fg_20071008cme}
\end{figure*}

This CME was initiated close to the western limb as seen from STEREO
B. Hereafter all the observations used are from instruments on board
the B spacecraft. Figure \ref{fg_20071008cme} shows five images of
the CME at different distances from the Sun. The CME was accompanied
by a prominence eruption starting at about 07:00 UT on October 8, as
seen by EUVI. The CME source region is clearly shown in the EUVI
304\AA\ image on the top-left panel of the figure. The erupting
prominence was also seen in the COR1 running-difference image (the
top-right panel). The CME was first observed in COR1 at about 08:46
UT on October 8, and continuously ran through COR2 and HI1 fields of
view (FOV). It even showed in the HI2 FOV after about 12:00 UT on
October 10. Since the CME was launched from the western limb and
showed a circular-like structure, we believe that the CME was viewed
by the instruments through an axial-view angle. Therefore, the
projection of the CME on the plane of the sky can be treated as the
cross-section of the CME.

To obtain the two quantities, $R$ and $L$, for necessary model inputs,
here we simply measure three parameters,
the heliocentric distance of the CME leading edge, $h$, and the maximum
and minimum position angles, $\theta_{max}$ and $\theta_{min}$, of the CME
as shown in Figure \ref{fg_coor}. Under the assumption of a circular
cross-section, $R$ and $L$ could be derived by
\begin{eqnarray}
R&=&h-L-R_S \\
L&=&\frac{h}{1+\sin \frac{\theta_{max}-\theta_{min}}{2}}-R_S
\end{eqnarray}
It should be noted that the measurements in HI2 images are not included
in the following analysis, because the elongation effect is not negligible.

\begin{figure*}[tbh]
  \centering
  \includegraphics[width=0.48\hsize]{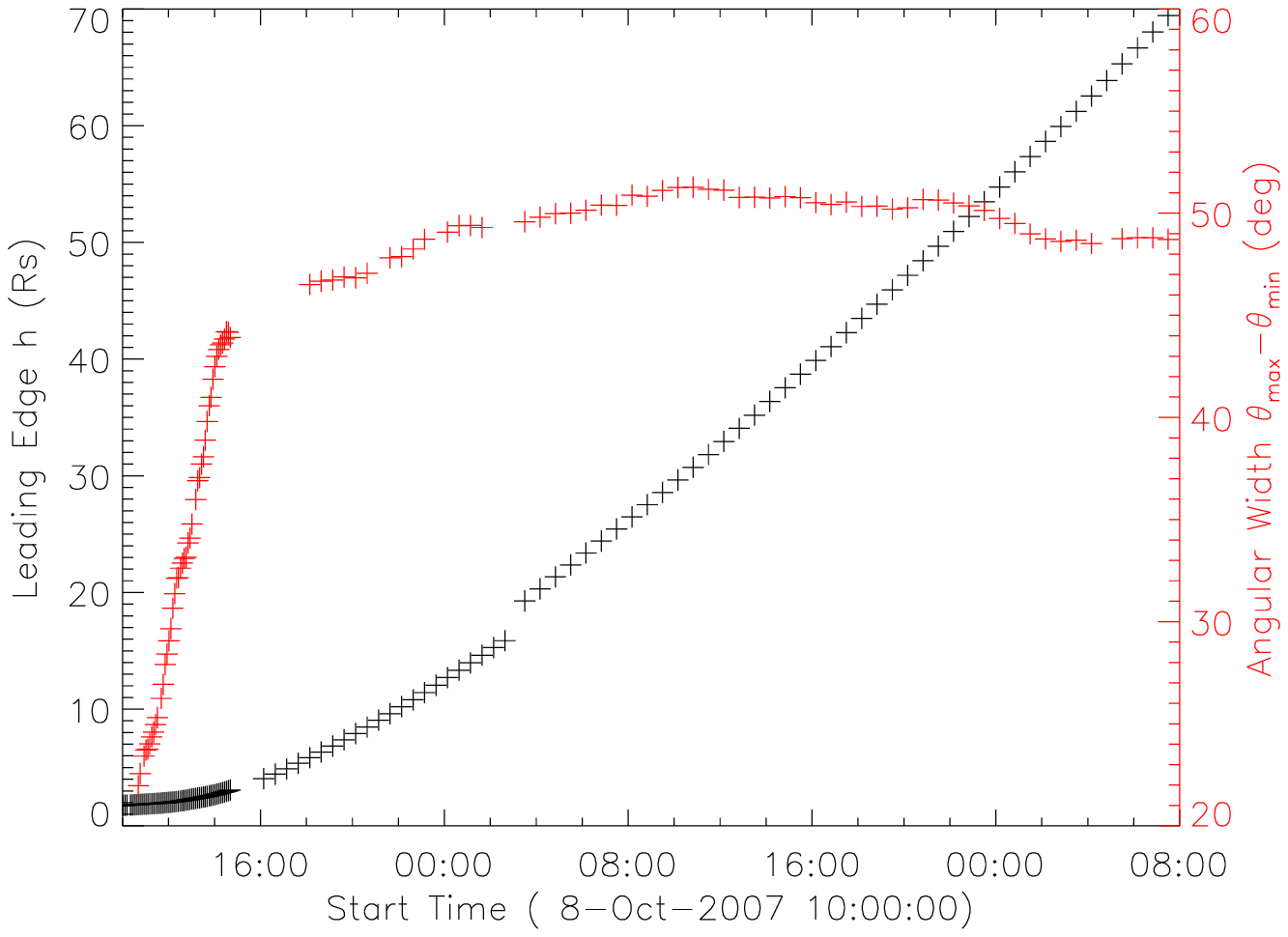}
  \includegraphics[width=0.45\hsize]{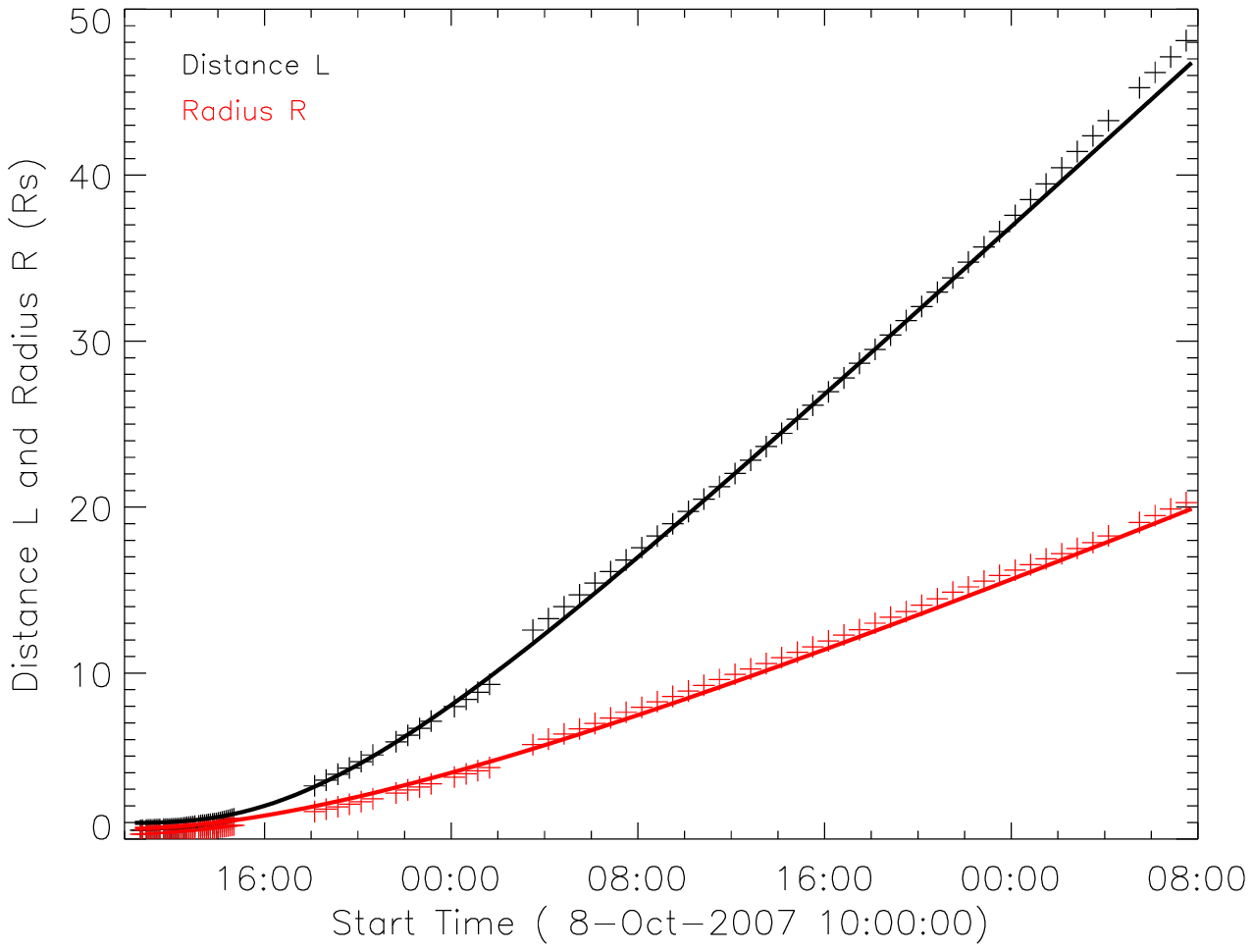}\\
  \hskip -170pt
  \includegraphics[width=0.45\hsize]{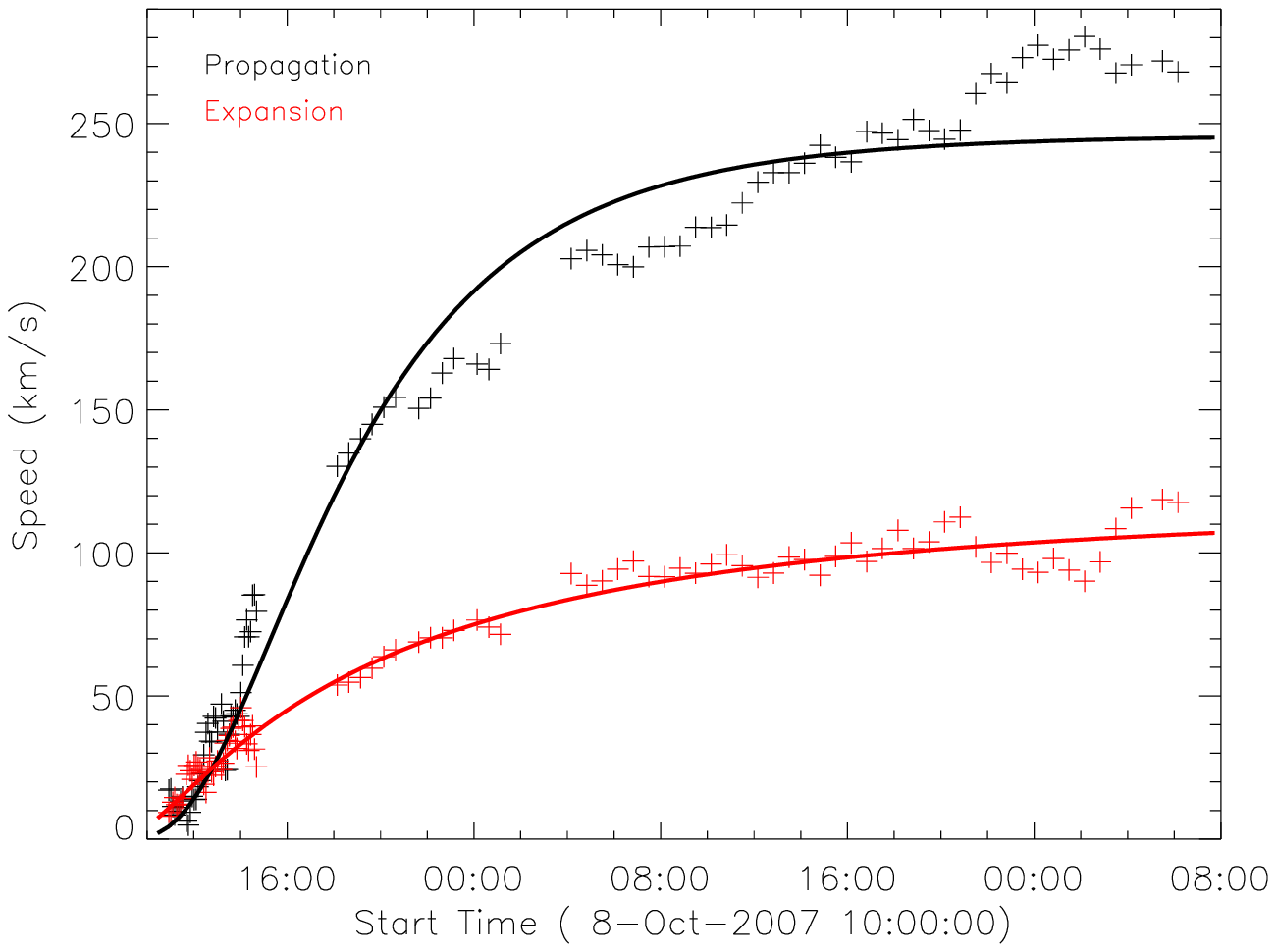}
  \caption{{\it Left-upper panel:} The measurements of the heliocentric distance,
$h$, of the flux-rope CME leading edge and its angular width,
$\Delta\theta=\theta_{max}-\theta_{min}$. {\it Right panel:} The
derived distance, $L$, of the flux rope axis from the solar surface
and the flux rope radius, $R$. {\it Left-lower panel:} The
propagation, $v_c$, and expansion, $v_e$, speeds derived from $L$
and $R$, respectively.}\label{fg_cme_motion}
\end{figure*}

Figure \ref{fg_cme_motion} shows the measurements and the derived parameters.
The CME is a slow and gradually accelerated event. It took about 46 hours
for its leading edge to reach 70 $R_S$. Nevertheless, because of its slowness,
we are able to make about one hundred measurement points for this CME. The
red crosses plotted in the leftupper panel suggests that the CME angular width
increased at the early phase (mainly in the COR1 FOV), and then reached to
a near-constant value in the COR2 and HI1 FOVs. The right panel presents
the evolution of the derived $R$ and $L$.
It is shown that the radius of the flux-rope CME is about 20 $R_S$ when it
propagated nearly 50 $R_S$ away from the Sun, which put the leading edge at
about 70 $R_S$. The left-lower panel exhibits the speeds derived from the $R$ and $L$,
namely expansion, $v_e$ and propagation, $v_c$, speeds, respectively. At the early
phase, the expansion speed was very close to the propagation speed.
In the later phase, the propagation speed increased more
quickly than the expansion speed. The increased difference between $v_c$ and
$v_e$ is probably because of (1) the enhanced drag force of the ambient solar wind,
which is fully formed in the outer corona and (2) the weakened pressure in the CME.
The issue of CME acceleration, which is as important as CME expansion, is not
addressed in our model presented in this paper.

In the measurements, the CME radius obtained is the one along the
latitudinal direction on the meridional plane. This radius would be
the same as the radius along the radial direction if the
cross-section is a perfect circle. However, the true cross-section
deviates from the perfect circle, and the deviation becomes larger
as the CME is further from the Sun
\citep[e.g.,][]{Riley_Crooker_2004}. The distortional stretching of
the cross-section is caused by the divergent radial expansion of the
background solar wind, which causes kinematic expansion of CMEs
along both the meridional and azimuthal directions, but not at all
along the radial direction. The CME expansion along the radial
direction is mostly driven by the dynamic effect, such as pressure
gradient forces, while the expansion along the other two directions
that lie on the spherical surface is caused by the combination of
the dynamic and kinematic effects. As a result, the overall
cross-section is a convex-outward ``pancake" shape
\citep{Riley_Crooker_2004}. Figure \ref{fg_20071008cme}f shows such
a distortion of the 2008 November 8 CME as observed in HI1 FOV; the
aspect ratio, defined by the ratio of the radius along the
meridional direction and that along the radial direction, is about
1.4 when the CME leading edge is at $\sim70 R_S$.

Due to this stretching effect, our measurements assuming a circular cross-section lead to the inaccuracy of the
measured parameters and the inferred parameters as well. In order to study the internal state of a CME, the radius
of the CME, $R$, should be the one along the radial direction, and it is apparently overestimated when the radius
along the meridional direction is adopted. The derived expansion speed of CME is thus larger than the true value.
Such simplified measurements would infer unrealistic parameters of CME at 1 AU.  For instance, the observed radius
of 20 $R_S$ of the CME at a distance of 50 $R_S$ from the Sun would imply a CME cross-section of 0.8 AU at 1 AU, which is
too larger to be true. The observed speeds  of $v_e$ and $v_c$ would imply a speed of about 150 km/s at the trailing
edge of the CME, which is much smaller than the observed solar wind speed, i.e., about 300 km/s. Therefore, one
should be cautious when our method is applied to CMEs at a large distance from the Sun (e.g, $> 70 R_S$). The
heliospheric region we investigated in this paper is within about 70 $R_S$, and the stretching effect is
relatively small. Nevertheless, we will carefully estimate the errors on CME parameters in the second
paragraph of Sec.\ref{sec_discussion}.
We point out here that there is an observational difficulty in measuring the radius of CMEs along the radial
direction in a consistent way, mainly because of the low brightness contrast of the CME trailing edge in coronagraph
images. This difficulty might be overcome if the CME of interest is particularly bright.

Before modeling the CME, we fit the measurement points with a certain function
to retrieve the smooth evolution process of the CME, which is required for the
model. We use the modified function of log normal distribution to fit the speeds.
We did not fit the expansion acceleration directly, because any small
error in measurements of $R$ will be dramatically amplified in its second derivative
$a_e$.
The fitting function of velocity has the form
\begin{eqnarray}
v(t)=\frac{v_\infty}{2}\left[1+\mathrm{erf}\left(\frac{\ln(t-t_0)-M}{S\sqrt{2}}\right)\right]
\end{eqnarray}
where $\mathrm{erf}(z)$ is the erf function or error function, defined by
\begin{eqnarray}
\mathrm{erf}(z)\equiv\frac{2}{\sqrt{\pi}}\int_0^ze^{-t^2}dt
\end{eqnarray}
This function has a value range from 0 to $v_\infty$. It is chosen
because the measurements show a trend that, at least within the FOVs of SECCHI,
both the speeds will not increase forever, but instead
asymptotically approach a constant speed, $v_\infty$. The acceleration
can be derived by
\begin{eqnarray}
a_e(t)=\frac{v_\infty}{S\sqrt{2\pi}(t-t_0)}e^{-\frac{[\ln(t-t_0)-M]^2}{2S^2}}
\end{eqnarray}
The solid lines in the left-lower panel of Figure
\ref{fg_cme_motion} show the fitting results. The fitted parameter
$v_\infty$ is 118 km/s for expansion and 246 km/s for propagation.
As a comparison with the measurements, the integrals of the fitting
curves of the speeds are also plotted in the right panel. It has
been mentioned before that these estimated speeds suffer the solar
wind stretching effect. Particularly, the estimated expansion speed
is larger than that it should be. The error will be discussed in
Sec.\ref{sec_discussion}.

\subsection{Results}\label{sec_results}
To fit the above curves with the model, Eq.\ref{eq_sase_fitting}, we use
an iterative method. Generally speaking, first we solve this equation
in every 8 neighboring measurement points to obtain a set of parameters $c_{0-6}$
and $\Gamma$. The segment of the 8 points is a running box through the entire
evolution process of the CME. Secondly, input the obtained variable $\Gamma$
into the model as guess values to fit the global constants $c_{0-6}$. Thirdly,
use the fitted $c_{0-6}$ to update the variable $\Gamma$ by solving
Eq.\ref{eq_sase_fitting} again. Then iterate the above 2nd and 3rd steps to
make constants $c_{0-6}$ and $\Gamma$ converging to a steady solution. For the sake
of simplicity, we ignore the poloidal motion of the fluid by setting $c_0$ zero. It
is also because there seems no strong observational evidence showing a ring flow
inside a CME.

The model results are shown in Figure \ref{fg_cme_state}.
The uncertainty of the model results is estimated from the relative error
of $a_e$, which is given by
\begin{eqnarray}
E=\left|\frac{a_{em}-a_{ei}}{a_{ei}}\right|
\end{eqnarray}
where $a_{ei}$ is the value calculated by the input data, and $a_{em}$ is
the model value. The error curve is plotted in the left-upper panel of Figure
\ref{fg_cme_state}. It is found that the error is
smaller than 1\%, except during 12:00 -- 18:00 UT.
A possible explanation of the large uncertainty during that time has been
given in the last second paragraph of Sec.\ref{sec_discussion}.

\subsubsection{Polytropic Index}
\begin{figure*}[tbh]
  \centering
\hskip -20pt  \includegraphics[width=0.45\hsize]{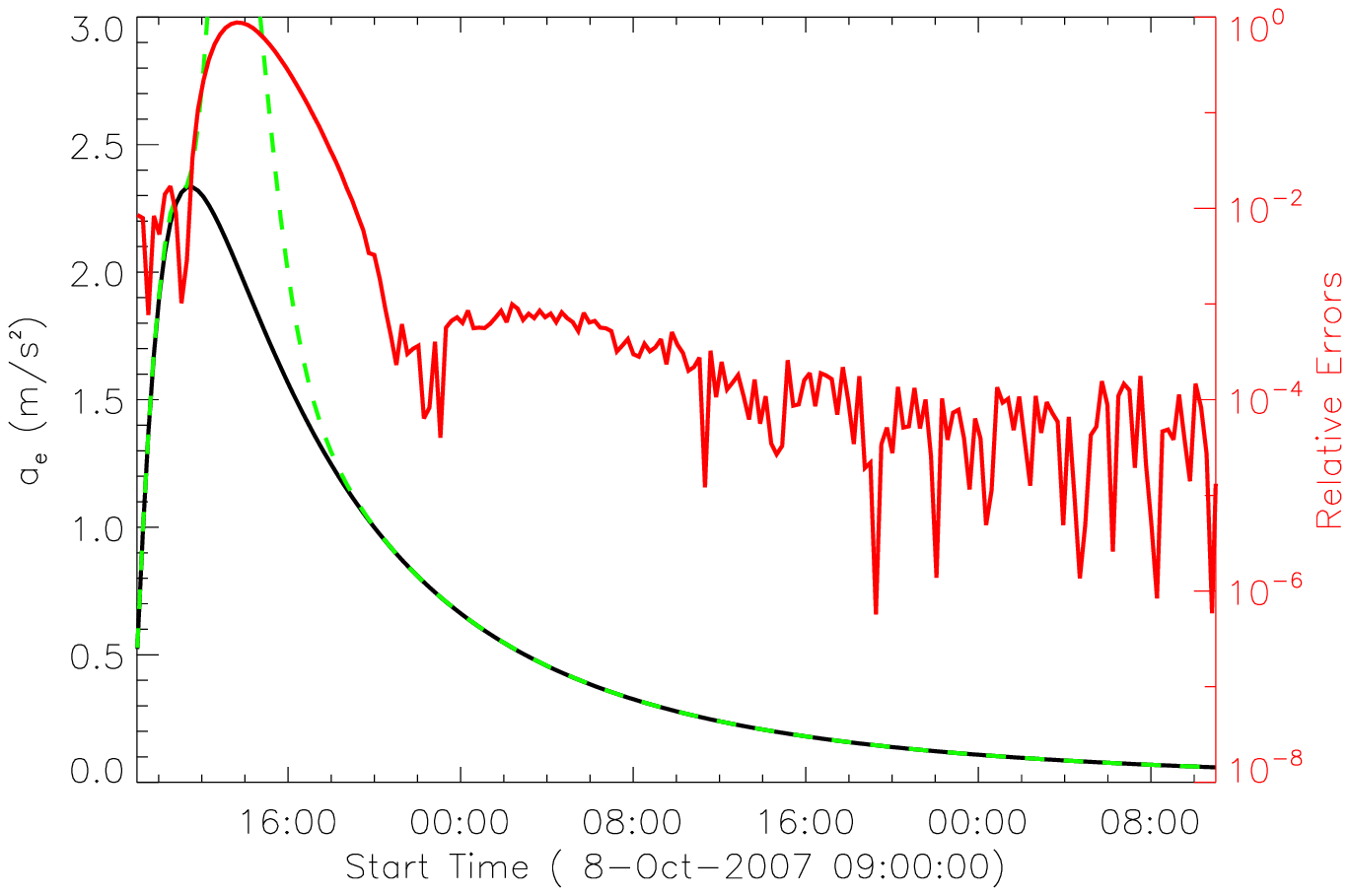}
\hskip 10pt  \includegraphics[width=0.41\hsize]{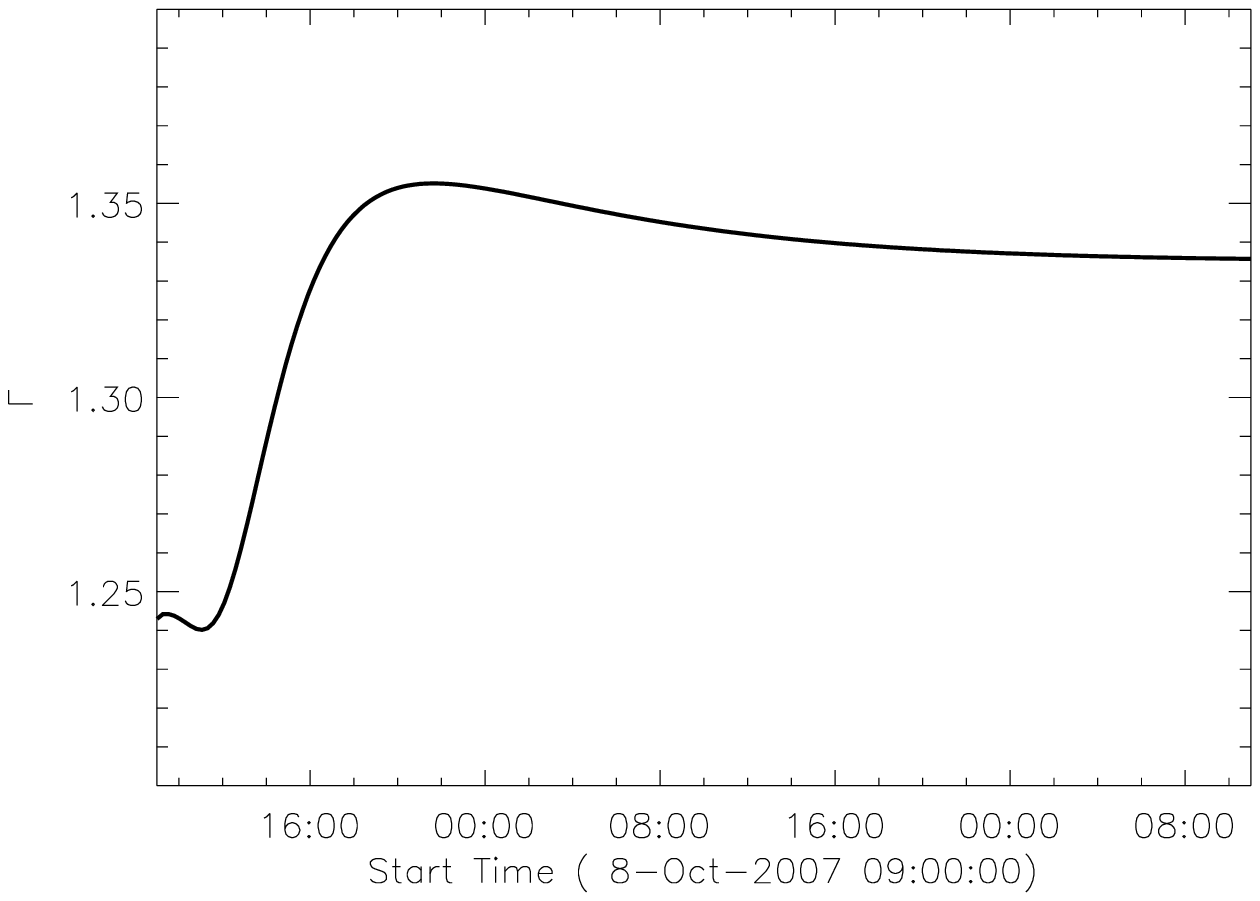}\\
\hskip 0pt  \includegraphics[width=0.415\hsize]{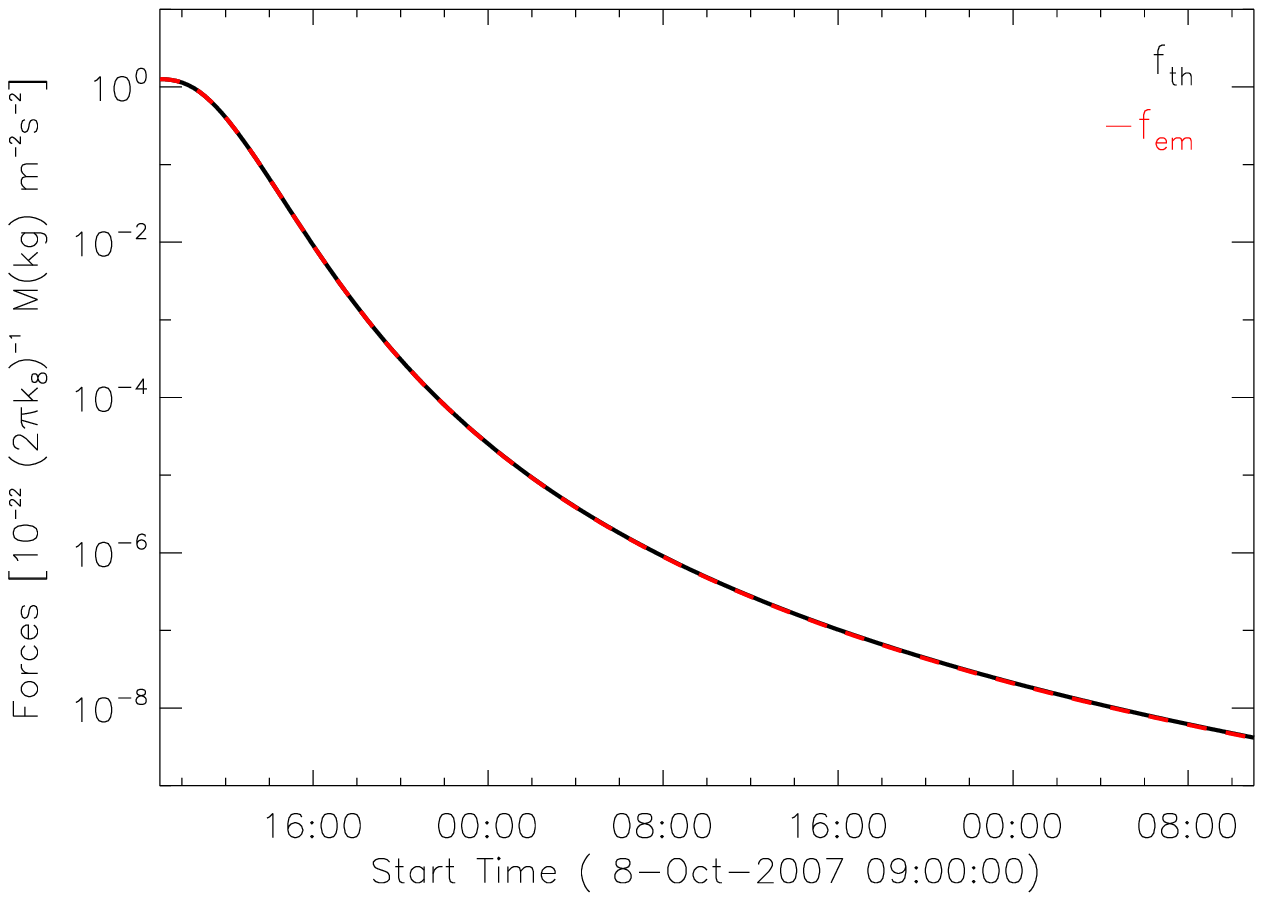} \hskip
10pt
\includegraphics[width=0.495\hsize]{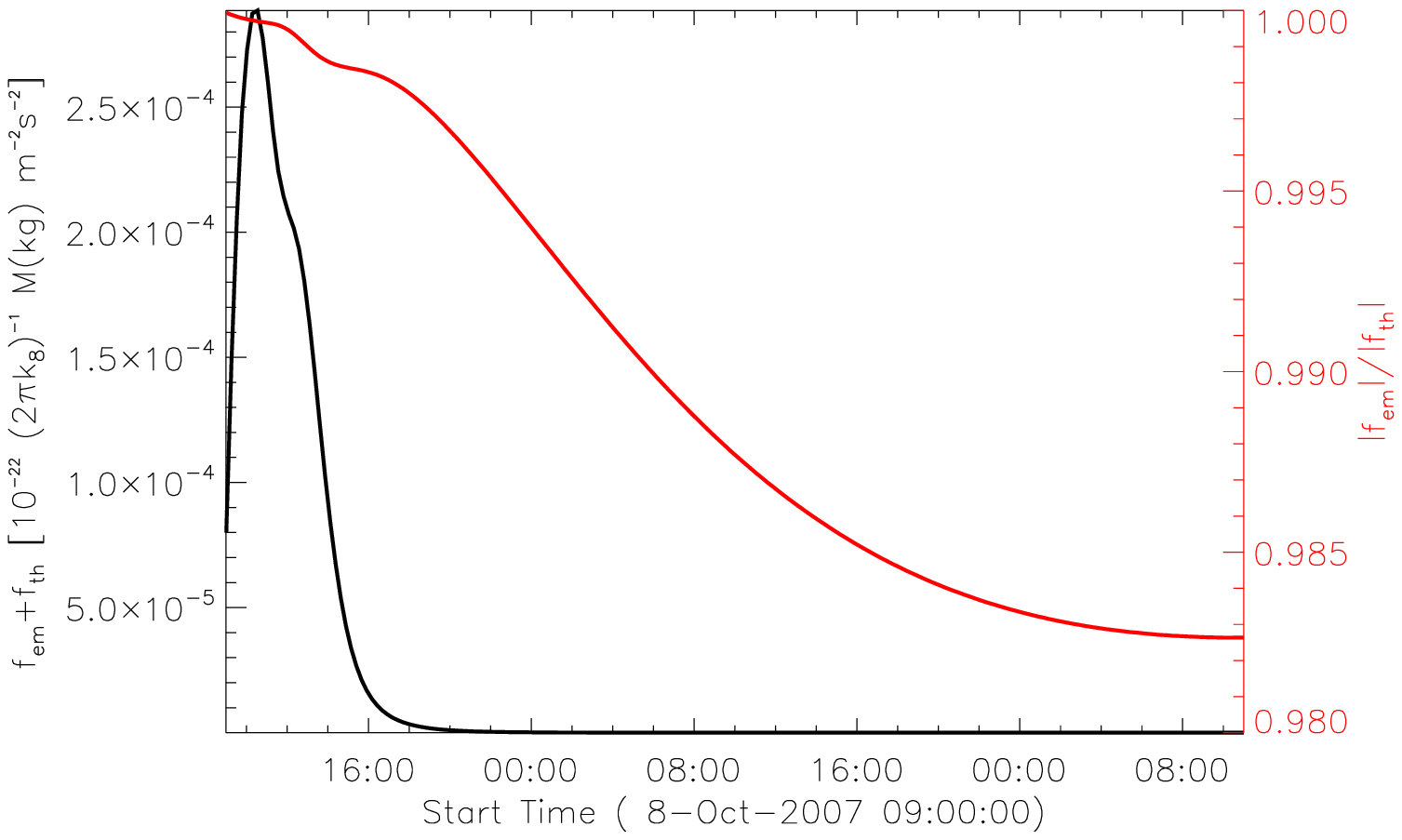}
  \caption{{\it Left-upper panel:} The profile of $a_e$ (black), the modeled result (dashed green),
and the relative error (see text for details). {\it Right-upper panel:} The variation of the polytropic index, $\Gamma$,
of the CME plasma. {\it Left-lower panel:} The variations of the average Lorentz force, $f_{em}$, and the average
thermal pressure force, $f_{th}$. Their signs have been marked on the upper right corner. {\it Right-lower panel:}
The sum and ratio of the two forces.}\label{fg_cme_state}
\end{figure*}

From the right-upper panel of Figure \ref{fg_cme_state}, it is found
that $\Gamma$ was less than 1.4 throughout the interplanetary space.
In the inner corona, say $L\lsim2R_S$, it was about 1.24. After
entering the outer corona, it quickly increased to above 1.35 at
$L\approx5 R_S$, and then slowly approached down to about 1.336,
which is very close to the first critical point $\frac{4}{3}$. This
value of $\Gamma$ is consistent with the observational value
obtained from \citet{Liu_etal_2006} statistics for protons. As the
CME kept expanding during its propagation in the FOVs, the
polytropic index less than $\frac{5}{3}$ means that there must be
some mechanisms to inject heat from somewhere into the CME plasma.
Although the CME plasma continuously got thermal energy, the proton
temperature may be still much lower than that in the ambient solar
wind, as revealed by many in-situ observations of MCs
\citep[e.g.,][]{Burlaga_etal_1981}.

We believe that the hot plasmas in the lower solar atmosphere is
probably a major heat source of CMEs in the interplanetary space. As shown
in Figure \ref{fg_coor}, a CME is believed to be a looped structure
with two ends rooted on the solar surface in a global scale.
Bidirectional electron streams are one of the evidence of it
\citep[e.g.][]{Farrugia_etal_1993b, Larson_etal_1997}. Thus it is
possible that heat is conducted from the bottom to CMEs. The ambient
solar wind with higher temperature might be an additional source
because the temperature difference between the two mediums is
significant. However, the cross-field diffusion of particles are
much more difficult than the motion parallel to magnetic field
lines, especially in a nearly force-free flux rope; the coefficient
ratio $\kappa_\perp/\kappa_\parallel$ of perpendicular to parallel
diffusion roughly locates in the range of $0.005 - 0.05$
\citep[e.g.,][]{Jokipii_etal_1995, Giacalone_Jokipii_1999,
Zank_etal_2004, Bieber_etal_2004}. Thus the contribution of the
ambient high-temperature solar wind should be very limited.

It is well known that the magnetic energy decreases as CMEs
propagate away from the Sun. According to our model, the total
magnetic energy is given by
\begin{eqnarray}
E_m=\frac{1}{2\mu_0}\int B^2rdrd\phi dz=\pi\mu_0^{-1}(k_9l^{-1}+k_{10}R^{-2}l)
\end{eqnarray}
where
\begin{eqnarray}
&k_9=\int_0^1(\frac{\partial f_z}{\partial x})^2xdx \\
&k_{10}=\int_0^1[\frac{\partial}{x\partial x}(xf_\phi)]^2xdx \label{eq_menergy}
\end{eqnarray}
are both positive integral constants. The magnetic energy generally
dissipates at the rate of $\sim l^{-1}$, which is a significant
dissipation as CMEs move outward. However, such magnetic energy
dissipation does not necessarily mean to be a major source of the
heat. According to MHD theory, magnetic energy partially goes into
kinetic energy and partially converts to thermal energy. The former
is due to the work done by Lorentz force ($\vec j\times\vec
B\cdot\vec u$), and the latter is through the Joule heating
($\frac{j^2}{\sigma}$) process, where $j$ is the current density and
$\sigma$ is the electrical conductivity. Since $\sigma$ usually has
a large value in interplanetary medium, without anomalous
resistivity, the magnetic energy does not have an efficient way to
be converted to thermal energy. However, there are possibly many
non-ideal processes, such as turbulence, but not accounted by MHD
theory. Thus we do not know whether the dissipated magnetic energy
is a major source of heating or not.

\subsubsection{Internal Forces}
The averaged Lorentz force, $f_{em}$, and thermal pressure force,
$f_{th}$, have been presented in the left-lower panel of Figure
\ref{fg_cme_state}. Their absolute values are very close to each
other, and both of them decreased continuously throughout the
interplanetary space. The signs of the two forces are opposite.
$f_{em}$ is negative indicating a centripetal force, whereas
$f_{th}$ is positive, centrifugal. This result suggests that the
thermal pressure force contributed to the CME expansion, but the
Lorentz force prevented the CME from expanding.

The difference between the two forces can be seen more clear from
the right-lower panel of Figure \ref{fg_cme_state}. The black line
exhibits the net force, $f_{em}+f_{th}$, inside the CME. It directed
outward and reached the maximum at about 10:30 UT. The profile is
consistent with the expansion acceleration presented in the
left-upper panel (the black line). Thus the net force just shows us
the internal cause of the CME expansion. The red line is the ratio
of their absolute values. Its value changed in a very small range
from about 1.0 to 0.98. It suggests that such a small difference
between the two forces is able to drive the CME expanding with the
acceleration at the order of 1 $m/s^2$. Moreover, the ratio decrease
means that the Lorentz force decreased slightly faster than the thermal
pressure force. One may notice that, since $\Gamma$ was larger than
the first critical point $\frac{4}{3}$ at $L\gsim5R_S$, according to the analysis in
Sec.\ref{sec_asymptotic}, the Lorentz force should drop slower than
thermal pressure force. Actually it may not be an inconsistency, because
the inference derived in Sec.\ref{sec_asymptotic} is established on
the force-free assumption, the CME we studied may not be
perfectly force-free, and therefore the first critical point of $\Gamma$ probably shifts a little bit.

Usually, CMEs are a flux rope with two ends rooted on the Sun. The
axial curvature of the flux rope may cause the magnetic strength at
the Sun-side of the flux rope larger than that at the opposite side,
which leads the Lorentz force having an additional component to
drive the flux rope moving outward away from the Sun
\citep[e.g.,][]{Garren_Chen_1994, Lin_etal_1998, Lin_etal_2002,
Kliem_Torok_2006, Fan_Gibson_2007}. Thus, as the flux rope we
applied here is assumed to be a straight cylinder, the Lorentz force
$f_{em}$ we derived does not include the component caused by the
axial curvature of the flux rope. This component is important in
studying the propagation properties of a CME. However, our model is
to study the CME internal state (specifically the thermodynamic process
and expansion behavior), and its propagation behavior is obtained
directly from coronagraph observations, thus the neglect of this
component should be acceptable although it does bring on some error,
which has been briefly mentioned in the
second paragraph of Sec.\ref{sec_discussion}.

\section{Summary} \label{sec_summary}
In this paper, we developed an analytical flux rope model for the
purpose of probing the internal state of CMEs and understanding its
expansion behavior. The model suggests that, if the flux rope
is force free, there are two critical
values for the polytropic index $\Gamma$. One is $\frac{4}{3}$, above/below
which the absolute value of the Lorentz force decreases slower/faster than
that of the thermal pressure force as the flux-rope CME propagates away
from the Sun. The other is $\frac{2}{3}$, above which the flux-rope CME
will essentially approach a steady expansion and propagation state.

By applying this model to the 2007 October 8 CME event, we find that
(1) the polytropic index of the CME plasma increased from initially 1.24
to more than 1.35 quickly, and then slowly decreased to about 1.336; it
suggests that there be continuously heat injected/converted into
the CME plasma and the value of $\Gamma$ tends to be the first critical value $\frac{4}{3}$;
(2) the Lorentz force directed inward while the thermal pressure force outward, both
of them decreased rapidly as the CME moved out, and the small difference between
them is consistent with the expansion acceleration of the CME; the direction of
the two forces reveal that the thermal pressure force is the internal driver of the CME expansion,
whereas the Lorentz force prevented the CME from expanding.

\section{Discussion}\label{sec_discussion}
In our model, the interaction between CMEs and the solar wind has
been implicitly included to certain extent, though we do not
explicitly address these effects. The consequences of the
interaction, in terms of the effects on the CME dynamic evolution
can be roughly classified into the following three types: (1) the
solar wind dragging effect, which is is due to the momentum exchange
between the CME plasma and the ambient solar wind and mainly affects
the CME's propagation speed or the bulk motion speed, (2) the solar
wind constraint effect on expansion, which is caused by the presence
of the external magnetic and thermal pressures and mainly prevents a
free expansion of the CME (i.e., in all directions), and (3) the
solar wind stretching effect on expansion, which is caused by the
divergent radial expansion of solar wind flow, and causes flattening
or ``pancaking" of CMEs. The first two effects are indirectly
included in the model through the measurements of $L$ and $R$.
Different dragging and/or constraint force(s) may result in
different variation of $L$ and/or $R$ with time (or heliocentric
distance). Particularly, we do not need to explicitly put the solar
wind dragging term in the model, because we are addressing the
internal state of CMEs, not the bulk acceleration. The stretching
effect, which is of a kinematic effect, is not included in our
model. As discussed in the sixth paragraph of
Sec.\ref{sec_measurements}, this is largely due to the limitation of
the measurements. The possible errors caused by such effect are
explicitly addressed in the next paragraph.

The main uncertainty of this model, we believe, comes from the
assumption of an axisymmetric cylinder, in which the curvature of
the axis of the flux rope and the distortion of the circular
cross-section are not taken into account. As to the first one, the
neglect of the axial curvature generally results in the Lorentz
force underestimated. As to the second one, as discussed earlier,
the distortion of the CME cross-section is due to the kinematic
stretching effect of a spherically divergent solar wind flow
\citep[e.g.,][]{Crooker_Intriligator_1996, Russell_Mulligan_2002,
Riley_etal_2003, Riley_Crooker_2004, Liu_etal_2006}. In the case of
the particular CME studied in this paper, the aspect ratio is about
1.4 when the CME leading edge is at $\sim70 R_S$  (or the flux rope
axis is at $\sim56 R_S$). The overall shape of the CME looks like an
ellipse. To estimate the errors caused by the circular assumption,
we approximate the ellipse to be a circle of the same area. With
this treatment, we estimate that $R$ is overestimated by 19\%, and
$L$ is underestimated by 11\%. Therefore, the expansion speed is
overestimated by 19\%, and the propagation speed is underestimated
by 11\%.  Further, we find that the density is underestimated by
21\% (ref to Eq.\ref{eq_rho_3}), $f_{th}$ is underestimated by 39\%
(ref to Eq.\ref{eq_pth} and assume $\Gamma=\frac{4}{3}$), $f_{em}$
is underestimated by 25--58\% (ref to Eq.\ref{eq_pem3}), and the
error of the polytropic index is probably neglected (ref. to
Eq.\ref{eq_fth}). These errors are evaluated for the CME at $\sim70
R_S$. At a smaller distance, we expect that the errors be smaller,
since the distortion is less severe.

The self-similar assumption made in our model may be another error
source, in which we assume that the distributions of the quantities
along $\hat{\vec r}$ in the flux rope remain unchanged during the
CME propagates away from the Sun. Self-similar evolution is a frequently
used assumption in modeling \citep[e.g.,][]{Low_1982, Kumar_Rust_1996,
Gibson_Low_1998, Krall_StCyr_2006}. The recent research by
\citet{Demoulin_Dasso_2009} suggested that, when $l$, the length of flux
rope, is proportional to $p_t^{-1/4}$, the total pressure in the ambient
solar wind, a force-free flux rope will evolve self-similarly. The
total pressure of solar wind consists of
thermal pressure $p_{th}=nkT$ and magnetic pressure
$p_{m}=\frac{B^2}{2\mu}$. Near the Sun, we can assume that the magnetic
pressure is dominant, thus it is approximated that $p_t\approx p_m\propto L^{-4}$,
i.e., $p_t^{-1/4}\propto L$. Since the length of a flux rope is
usually proportional to the distance $L$, we have
$l\propto p_t^{-1/4}$. It means that self-similar assumption should
be a good approximation when the CME is nearly force-free and not too
far away from the Sun. Other previous studies also showed that the
self-similar evolution of CMEs is probably true within tens solar radii
\citep[e.g.,][]{Chen_etal_1997, Krall_etal_2001, Maricic_etal_2004}.
On the other hand, however, the self-similar assumption must be broken
gradually. An obvious evidence is from the solar wind stretching effect
as have been addressed before. Another evidence is that a CME may relax
from a complex structure to a nearly force-free flux rope structure, for
example the simulation by \citet{Lynch_etal_2004}.

For the CME plasma, neglecting the viscous stress tensor
in Eq.\ref{eq_mom_1} might be appropriate. The viscous stress tensor
of protons can be approximately given by
\begin{eqnarray}
S_{ij}\approx3\eta_0\left(\frac{\delta_{ij}}{3}-\frac{B_iB_j}{B^2}\right)\left(\frac{\vec{B}\cdot\vec{B}\cdot\nabla\vec{v}}{B^2}-\frac{\nabla\cdot\vec{v}}{3}\right)
\end{eqnarray}
and $\eta_0$ is the coefficient of viscosity that could be estimated
by $\eta_0\approx10^{-17}T_p^{\frac{5}{2}}$ kg$\cdot$m$^{-1}\cdot$s$^{-1}$
\citep{Braginskii_1965, Hollweg_1985}.
Here $\delta_{ij}$ is the unit tensor, $\vec{v}$ is flow velocity, and
$T_p$ is the proton temperature.
Since the proton temperature in CMEs is low, $\eta_0$ and therefore
the viscous stress tensor is very small. Thus, we guess that the
viscosity in the momentum equation might be ignored.

Both forces ignored in Eq.\ref{eq_mom_1}, the gravity $\vec{F}_g$ and the equivalent fictitious force $\vec{F}_a$
due to the use of a non-inertial reference frame, are in the  radial direction in the solar frame.
Their effects can be evaluated by comparing them with the acceleration of
the expansion of the fluids in flux rope CMEs. The solar
gravity acceleration is about 270 m/s$^2$ at the surface, and decreases
at the rate of $r^{-2}$, which makes it as low as $\sim$2.7 m/s$^2$
at 10 $R_S$. And $\vec{F}_a$ should be also very small for most CMEs beyond
10 $R_S$. Thus both forces would significantly distort the model results
only on CMEs with slow expansion acceleration in the lower corona, but not on
those with large expansion acceleration or in the outer corona.
This may be the reason why a large error of $a_e$ appears during 12:00 -- 18:00 UT
in modeling this CME (left-upper panel of Fig.\ref{fg_cme_state}).

The flux-rope model presented in this paper might be the first of its kind
to provide a way to infer the inter state of CMEs directly based on coronagraph
observations. It is different from other CME dynamic models, such as those
by \citet{Chen_1989} and \citet{Gibson_Low_1998}, which were designed to study
the interaction of CMEs with the ambient solar wind and other dynamic processes
by adjusting the initial conditions of CMEs and the global parameters of the
ambient solar wind. Besides, \citet{Kumar_Rust_1996} proposed a current-core
flux rope model with self-similar evolution (ref. to KR model thereafter).
Although a self-similar flux rope
is also employed in their model, our model is largely different from theirs.
First, the flux rope in KR model is assumed force-free and the Lundquist
solution \citep{Lundquist_1950} is applied to describe the internal
magnetic structure, but our model does not specify the magnetic field
distribution and it may be non-force-free. Secondly, the self-similar assumption
in KR model limits the radius of the flux rope to be proportional to the
distance, whereas our self-similar condition is held only in the cross-section
of the flux rope; the $R$ and $L$ in our model are two independent measurements
(see Fig.\ref{fg_cme_motion}). Thirdly, KR model did not consider the solar wind
effects on the flux rope, while two of three solar wind effects are implicitly
included in our model. Thus one can treat our model a more generic one.
Undoubtedly, KR model is an excellent model for force-free
flux ropes, and got many interesting results. For example, it is suggested that the
polytropic index is $\frac{4}{3}$ for a CME far from the Sun. It is an inference
from their self-similar assumption, and it seems to be true for the 2007 October
CME we studied here. In our model, the $\Gamma$ value of $\frac{4}{3}$ implies a
special case (Sec.\ref{sec_asymptotic}) in which the two internal forces $f_{em}$
and $f_{th}$ vary at the same rate. Further work will be performed to test whether
it holds for all CME events.

\paragraph{Acknowledgments.}
We acknowledge the use of the data from
STEREO/SECCHI. We are grateful to James Chen and Yong C.-M. Liu for
discussions. We also thank the referees for valuable comments.
Y. Wang and J. Zhang are supported by grants from
NASA NNG05GG19G, NNG07AO72G, and NSF ATM-0748003. Y. Wang and C. Shen
also acknowledge the support of China grants from NSF 40525014, 973
key project 2006CB806304, and Ministry of Education 200530.

\section*{Appendix}
In cylindrical coordinate system, the magnetic field of a force-free
flux rope has the \citet{Lundquist_1950} solution
\begin{eqnarray}
B_r&=&0 \nonumber\\
B_\phi&=&HB_0J_1(2.41x) \\
B_z&=&B_0J_0(2.41x) \nonumber
\end{eqnarray}
where $x=\frac{r}{R}$ is the normalized radial distance as defined in Sec.\ref{sec_model},
$J_0$ and $J_1$ are the zero and first order Bessel functions, $H=\pm1$
indicates the sign of the handedness and $B_0$ is the magnetic field
magnitude at the axis of the flux rope. According to the properties
of Bessel function, we have the magnetic vector potential
\begin{eqnarray}
A_\phi&=&\frac{RB_0}{2.41}J_1(2.41x) \label{eq_ffaphi}\\
A_z&=&\frac{HRB_0}{2.41}J_0(2.41x) \label{eq_ffaz}
\end{eqnarray}
The conservation of $\Phi_z$
\begin{eqnarray}
\Phi_z=2\pi\int_0^R\frac{\partial}{\partial r}(rA_\phi)dr=\frac{2\pi R^2B_0}{2.41}J_1(2.41)=\mathrm{constant}
\end{eqnarray}
requires that
\begin{eqnarray}
B_0=2.41a_1R^{-2} \label{eq_ffb0}
\end{eqnarray}
where $a_1$ is a constant. The magnetic vector potential can be
rewritten as
\begin{eqnarray}
A_\phi&=&\frac{a_1}{R}J_1(2.41x) \label{eq_ffA_phi}\\
A_z&=&\frac{a_1H}{R}J_0(2.41x) \label{eq_ffA_z}
\end{eqnarray}
Meanwhile, the magnetic helicity is
\begin{eqnarray}
H_m=\int\vec{B}\cdot\vec{A}drrd\phi dz=4.82\pi a_1^2a_2HR^{-1}l
\end{eqnarray}
where $a_2=\int_0^1x(J_0^2+J_1^2)dx$ is a constant.
The conservation of $H_m$ results in
\begin{eqnarray}
&R\propto l
\end{eqnarray}
Combined it with the assumption Eq.\ref{eq_ll}, it is inferred that
\begin{eqnarray}
&R\propto L
\end{eqnarray}
which means that the force-free flux rope expands radially.

\bibliographystyle{agufull08}
\bibliography{../../ahareference}

\end{document}